\documentclass[referee]{raa}            

\usepackage{graphicx,times}             
\usepackage{amsmath,amssymb,bm}
\usepackage{multirow, multicol}
\usepackage{url}             
\usepackage{natbib}

\bibpunct{(}{)}{;}{a}{}{,}
\usepackage[a4paper=true]{hyperref}
\hypersetup{colorlinks = true, linkcolor = blue, anchorcolor = red, citecolor = blue, filecolor = red, pagecolor = red, urlcolor = red}

\begin{document}
   \title{The Magnification Invariant of Circularly-symmetric Lens Models}

   \volnopage{Vol.0 (20xx) No.0, 000--000}      
   \setcounter{page}{1}          

   \author{Chengliang Wei
      \inst{1,2}
   \and Zhe Chu
      \inst{1,3}
   \and Yiping Shu
      \inst{1,3}
   }

   \institute{Purple Mountain Observatory, the Partner Group of MPI f\"ur Astronomie, 8 Yuanhua Road, Nanjing 210033, China;
               {\it chengliangwei@pmo.ac.cn}\\
        \and
              University of Chinese Academy of Science, 19A Yuquan Road, 100049 Beijing, China \\
        \and
              School of Astronomy and Space Sciences, University of Science and Technology of China, 230029 Hefei, China 
             }

   \date{Received~~2009 month day; accepted~~2009~~month day}

\abstract{
In the context of strong gravitational lensing, the magnification of image is
of crucial importance to constrain   various lens models. For
several commonly used quadruple lens models, the magnification invariants,
defined as the sum of the signed magnifications of images, have been
analytically derived when the image multiplicity is a maximum. In this
paper, we further study the magnification of several disk lens models, including
(a) exponential disk lens, (b) Gaussian disk lens, (c) modified Hubble profile
lens, and another two of the popular three-dimensional symmetrical lens model,
(d) NFW lens and (e) Einasto lens. We find that magnification invariant does
also exist for each lens model. Moreover, our results show that magnification
invariants can be significantly changed by the characteristic surface mass
density $\kappa_{\rm c}$.
\keywords{Gravitational lensing: strong -- Methods: numerical}
}

\authorrunning{C. Wei et al.}            
\titlerunning{Magnification invariants of circular lenses}  
\maketitle

\section{Introduction}\label{sec:introduction}
As one of the promising ways to explore cosmological information,
strong gravitational lensing has been investigated in many astrophysical
studies \citep[e.g., ][]{2009PhDT.......238P, 2015ApJ...806..185C, 2015aska.confE..84M, 2015MNRAS.452.2423Y}:
for determining the mass density profile of galaxies
\citep[e.g., ][]{2009ApJ...703L..51K, 2012Natur.481..341V, 2016ApJ...833..264S, 2017ApJ...851...48S},
measuring the masses of central black holes in distant quiescent galaxies \citep[e.g., ][]{2001MNRAS.323..301M, 2005ApJ...627L..93R},
and estimating Hubble constant using time delays between multiple images in the
observed lensing systems
\citep[e.g., ][]{1999MNRAS.304..349B, 2002ApJ...581..823F, 2007waas.work..127T, 2011PhRvD..84l3529L, 2014ApJ...788L..35S}.

In the context of strong gravitational lensing, multiple images of a given background source
can be produced when the lensing system has a well alignment
between the observer, the lens, and the source \citep{1981ApJ...244L...1B, 1980ApJ...238L..67D, 2006glsw.conf.....M}.
Based on the magnification theorem, the total image number of a lens
is odd if the lens has a smooth surface mass density. On the other hand, the gravitational
lensing is able to magnify the apparent brightness of the source \citep{1992grle.book.....S}.
Thus the connection between magnification and multiple images
is crucial in a variety of studies of gravitational lensing
\citep{2000MNRAS.311..689W, 2009JMP....50c2501A, 2009JMP....50h2504W, 2010GReGr..42.2011P, 2013PhRvD..87b4024T, 2015MNRAS.449.2079C}.
For example, the summation of the image magnification for  a point
lens and singular isothermal sphere (SIS) lens is 1 and 2 respectively,
regardless of the source positions and how large the Einstein radii are.
In \citet{1998ApJ...509L..13D}, the author found that the summation
of signed magnifications of all the images ($I = \Sigma_{i}\mu_i$, where
$\mu_i$ denotes the signed magnification of the $i^{\rm th}$ image) is a
constant for several quadrupole lens models when the image number is a maximum.
Conventionally, this constant ($I$) is dubbed 'magnification invariant'. In \citet{2015MNRAS.449.2079C},
they studied several lens models, including (1) singular isothermal elliptical
density (SIED), (2) singular isothermal elliptical potential (SIEP), (3) singular isothermal
quadrupole (SIQ), (4) SIS + external shear and (5) point + external shear lenses. The
magnification invariants for these models are shown in Tab.~\ref{tab:magrev}. It is
found that the invariant is independent of most of the model parameters, as long as
the source lies inside of the caustic.

\begin{table*}
	\renewcommand\arraystretch{1.5}
	\centering
	\caption{Magnification invariants for some four-image lens models$^{a}$.}
	\label{tab:magrev}
	\begin{tabular}{lll} 
		\hline
		Lens model  & Lens potential $\psi$ & $I^{b}$ $\left( =\Sigma_{i}\mu_i \right)$ \\
        \hline
        1. SIED($x, y$)        & $x\alpha_{x}+y\alpha_{y}$      &  $\approx 2.8$ \\
        2. SIEP($x, y$)        & $b\sqrt{q^{2}x^{2} + y^{2}}$   &  2          \\
        3. SIQ($\theta, \phi$)         & $\theta_{\rm E}\theta - \frac{1}{3}\theta_{\rm E}k\theta\cos2\phi$  &  1  \\
        4. SIS+shear($\theta, \phi$)   & $\theta_{\rm E}\theta - \frac{\gamma}{2}\theta^{2}\cos2\phi$        & $2/(1-\gamma^{2})$ \\
        5. point+shear($\theta, \phi$) & $\theta_{\rm E}^{2}\ln\theta - \frac{\gamma}{2}\theta^{2}\cos2\phi$        & $1/(1-\gamma^{2})$ \\
		\hline
	\end{tabular}
    \\
    \begin{flushleft}
    {\it Notes.} $^{a}$ Here $b$ is a constant parameter, $q$ is the axial ratio in the SIEP lens,
    $\theta_{\rm E}$ is the Einstein radius, $k(0 \leq k \leq 1)$ is the intensity of
    the quadrupole relative to the monopole in the SIQ lens, and $\gamma$ indicates
    the external shear \citet{2015MNRAS.449.2079C}.
    $^b$ \citet{1998ApJ...509L..13D, 2001JMP....42.1818D, 2000MNRAS.311..689W}
    \end{flushleft}
\end{table*}

If the surface mass density of a lens is circularly symmetric,
the lensing properties, such as deflection angle and lensing
shear, can be derived analytically. While it is not clear whether
a magnification invariant can be found in the circularly-symmetric
lens system (hereafter circular lens). In this paper, we study the
magnification invariant of the circular lens by considering several
commonly used lens models, like exponential disk lens and Gaussian
disk lens. Moreover, as found in the numerical $N$-body simulations,
the density profile of a cold dark matter halo can be described by
the NFW profile \citep{1997ApJ...490..493N}. While recent
high-resolution simulations show that Einasto radial profile, which
is a non-singular three-parameter model, provides a more accurate
description of dark matter haloes \citep{2007JCAP...07..006E, 2004MNRAS.349.1039N, 2010MNRAS.402...21N}.
For these two popular spherical  symmetrical density profiles,
we can derive their projected surface mass density by integrating
the three-dimensional density profile along the line of sight. The
investigation of the magnification invariant is important to constrain
these lens models.

The rest of this paper is organized as follows. In Sec.~\ref{sec:cirLens},
we briefly summarize the basics of circular lens, and several commonly
used lens models. In Sec.~\ref{sec:method}, we describe our method to derive
the magnification invariants for each discussed lens model and show the
dependency between derived invariants and lensing parameters. Conclusions and
discussions are summarized in Sec.~\ref{sec:summary}.

\section{CIRCULAR LENS}\label{sec:cirLens}

An arbitrary surface mass density $\Sigma(\bm{\theta})$ can be expanded in
terms of a complete set of orthogonal basis functions \citep[e.g., ][]{2000ApJ...535..671T},
which can decompose $\Sigma(\bm{\theta})$ into multipole components,
\begin{equation} \label{equ:me}
  \Sigma(\bm{\theta}) = \Sigma_{0}(\theta) + \sum_{m=1}^{\infty} \left[ {\rm A}_{m}(\theta)\cos(m\chi) + {\rm B}_{m}(\theta) \sin(m\chi) \right],
\end{equation}
where $\theta$ is the distance from the lens centre, $\chi$ is the polar angle.
The first term $\Sigma_0(\theta)=1/2\pi \int_{0}^{2\pi} \Sigma(\bm{\theta}){\rm d}\chi$
is known as the monopole. The higher-order multipoles ($m\geqslant1$) represent
the angular structure of the mass distribution, which are composed of two parts
(${\rm A}_{m}$ and ${\rm B}_{m}$). Because of the symmetry, circular lenses can be fully
described by its monopole. As a kind of ideal lens model, most of properties can be
given analytically for a given circular lens \citep{1991ApJ...370....1M, 1992grle.book.....S, 1999fsu..conf..360N}.
In this section, we present the basics  for our analysis,
and also describe several commonly discussed circular lens models. Throughout the
paper we denote the observed angular position on the lens plane by
$\bm{\theta}=\theta(\cos\chi,\sin\chi)$ and its source position by $\bm{\beta}$.

\subsection{Basics of Circular Lens}
In general, if the mass distribution of a lens object is circularly-symmetric,
the 2D Poisson equation, $\nabla^2\psi(\bm{\theta})=2\kappa(\bm{\theta})$,
can be reduced to a function only of the distance from the lens centre
$\theta=|\bm{\theta}|$,
\begin{equation} \label{equ:poisson}
  \frac{1}{\theta}\frac{\partial}{\partial\theta}(\theta\frac{\partial}{\partial\theta})\psi(\theta)=2\kappa(\theta),
\end{equation}
where $\psi(\theta)$ is the deflection potential, and $\kappa(\theta)=\Sigma(\theta)/\Sigma_{\rm crit}$
is the dimensionless surface mass density in unit of the critical surface mass density
$\Sigma_{\rm crit}={\rm c}^2 D_{\rm s}/4\pi G D_{\rm d} D_{\rm ds}$. Here $D_{\rm d}$,
$D_{\rm s}$ and $D_{\rm ds}$ are the angular diameter distance to the lens, to the
source and between the lens and the source, respectively. For circular lens,
deflection angle $\alpha(\theta)$, lensing shear $\gamma(\theta)$ and
magnification $\mu(\theta)$ are given as
\citep{1991ApJ...370....1M, 2006glsw.conf.....M},
\begin{align}
  & \alpha(\theta)=\theta\bar{\kappa}(\theta), \label{equ:alpha1} \\
  & \gamma(\theta)=\bar{\kappa}(\theta)-\kappa(\theta)=\frac{\alpha(\theta)}{\theta}-\kappa(\theta), \label{equ:gamma1} \\
  & \frac{1}{\mu(\theta)}=[1-\frac{\alpha(\theta)}{\theta}][1+\frac{\alpha(\theta)}{\theta}-2\kappa(\theta)]. \label{equ:mu1} 
\end{align}
where $\bar{\kappa}(\theta)=\bar{\Sigma}(\theta)/\Sigma_{\rm crit}$ is the
mean surface mass density inside $\theta$.
Since the critical curves arise at $1/\mu=0$, Eq.~\ref{equ:mu1} implies that
the circular lens has a pair of critical curves \citep{2006glsw.conf.....M}.
The one, $1-\alpha(\theta)/\theta=0$, is the tangential critical curve, which
corresponds to the Einstein Ring with Einstein radius. The another,
$1+\alpha(\theta)/\theta-2\kappa(\theta)=0$, is called radial critical curve,
which also defines a ring and the corresponding radius.

In the strong gravitational lensing, multiple images can be
produced for a given source. The image number depends on the position of the
source with respect to the caustics, which are the mapping of the critical curves
in the source plane. For circular lenses, since the tangential critical curve
does not lead to a caustic curve and the corresponding caustic degenerates to
a single point $\bm{\beta}=0$, the tangential critical curve has no influence
on the image multiplicity. Thus, pairs of images can only be created or destroyed
if the radial critical curve exists. When a source lies inside of the radial caustic,
three images will be produced at most. In general, to derive the
magnifications of the images, the lens equation should be solved numerically.

In the rest of this section, we will discuss several often used circular lens
models,  which can be analyzed analytically and are useful for theoretical investigations.

\subsection{Circular Lens Models} \label{ssec:simpleLens}
\begin{figure}
    \centering
    \includegraphics[width=0.75\columnwidth]{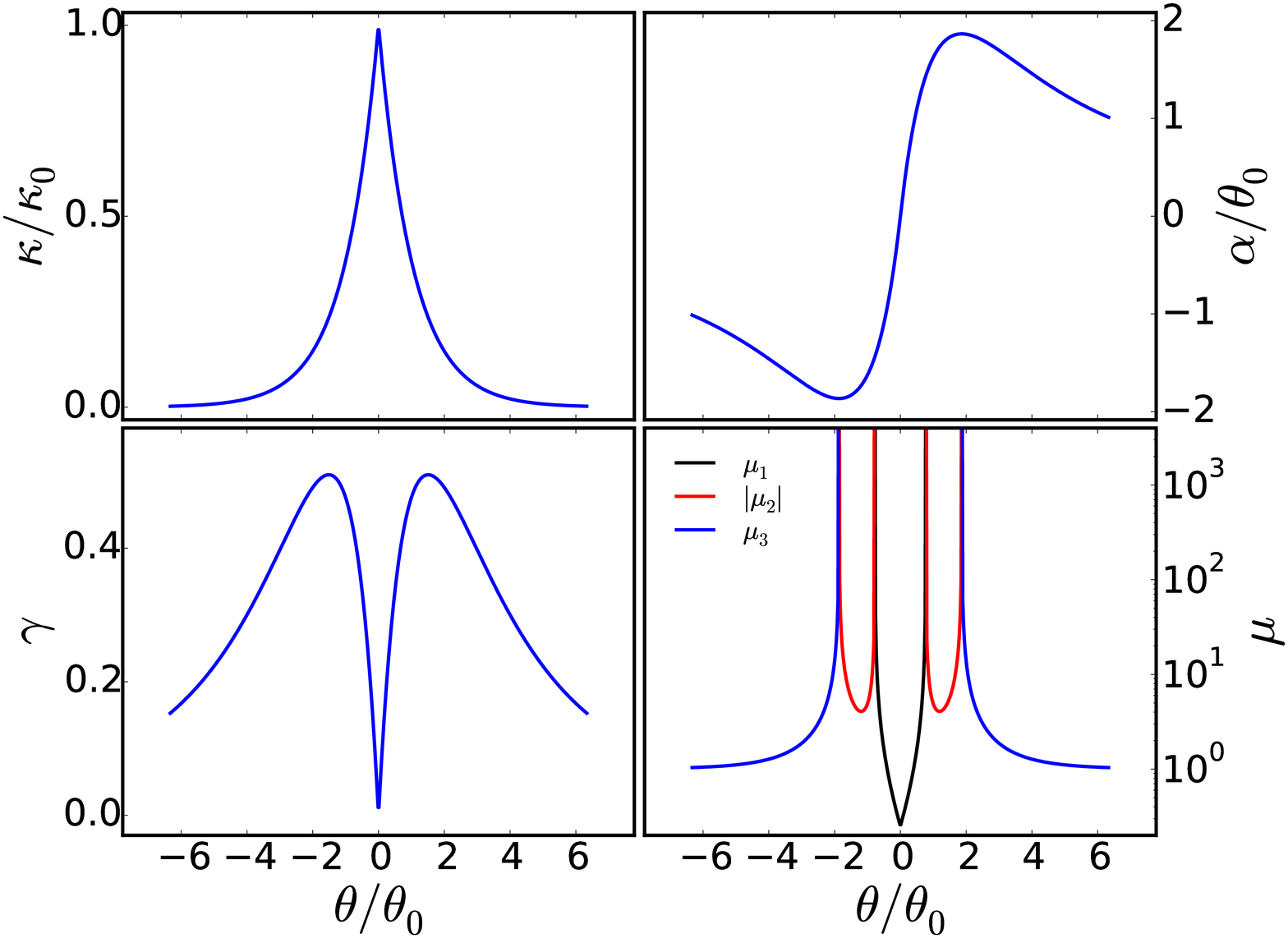}
    \caption{The convergence $\kappa(\theta)$, deflection angle
             $\alpha(\theta)$, shear $\gamma(\theta)$, and magnification
             $\mu(\theta)$ for a typical exponential disk lens.}
    \label{fig:expQua}
\end{figure}

The first lens model considered in this work is an exponential disk model,
usually used to describe the mass distribution of a spiral galaxy \citep{2000ApJ...536..584L}.
For an exponential disk model with $\Sigma(\theta)=\Sigma_0 {\rm exp}(-{\theta}/{\theta_0})$,
the convergence can be obtained as
\begin{equation} \label{equ:profileExp}
  \kappa_{{\rm E}}(\theta)=\kappa_0 {\rm exp}(-{\theta}/{\theta_0}).
\end{equation}
where $\kappa_{0}$ is the central dimensionless surface mass density and
$\theta_{0}$ is the scale length of the lens model. The scaled deflection angle
$\alpha(\theta)$ and lensing shear $\gamma(\theta)$ are derived as shown in Tab.~\ref{tab:table2},
and these lensing properties are shown in Fig.~\ref{fig:expQua}.
Clearly, two critical curves can be identified, where $\mu(\theta)=\infty$,
in the figure of the magnification (in the lower-right panel). Different colors indicate that lens plane
can be divided into three image region by the two critical curves. The Fermat
maximum image lies in the black line region around the centre of the lens, and it
has a positive magnification. The minimum image arises in the blue line region,
which is outskirt of the lensing disk, and its magnification is also positive.
The saddle image can be found in red line region, which has the negative magnification,
and here we flip the signs in the figure. In the left panel of Fig.~\ref{fig:IMAGEZONE},
we show the distribution of the absolute value of the magnification in logarithm
($\log|\mu|$) in the lens plane.

Based on the lens equation, the positions of images for a given source are calculated as
\begin{equation} \label{equ:leExp}
  \beta=\theta-\frac{2\kappa_0}{\theta} \left[ \theta^2_0-\theta_0(\theta+\theta_0){\rm exp}({-{\theta}/{\theta_0}}) \right].
\end{equation}
Here, the image position $\theta$ can not be solved analytically. Thereby, it is also
difficult to derive the magnifications of the three images. In this work, we use of a precise
numerical ray-tracing method, developed by \citet{2016MNRAS.461.4466C}, to calculate
the positions and corresponding magnifications of different images for a given source.
The results can be found in section \ref{sec:method}.

We also consider another two circular lens models:
The dimensionless density profile is given as
$\kappa_{{\rm G}}(\theta)=\kappa_0 {\rm exp}(-{\theta^2}/{\theta^2_0})$,
dubbed as Gaussian disk lens. This model can be used to describe Einasto lens
with index $1/2$, as represented in Sec.~\ref{sec:EinastoLens}. The other is
the so-called modified Hubble profile \citep{1972ApJ...175..627R}, and the
projected surface mass density of this lens model is given as,
\begin{equation} \label{equ:profileHubble}
    \Sigma(\theta)=\frac{\Sigma_0}{1+\theta^{2}/\theta_{0}^{2}},
\end{equation}
which is a softened power law lens, and the dimensionless projected
surface density can be expressed as $\kappa_{{\rm MH}}(\theta)=\kappa_0{\theta_0^{2}}/{(\theta^{2}+\theta_0^{2})}$.
For these two important lens models, their lensing properties ($\alpha$, $\gamma$,
and $\mu$) are listed in Tab.~\ref{tab:table2}, respectively.

\begin{table*}
	\renewcommand\arraystretch{1.5}
	\centering
	\caption{Lens properties of the three circular lens models.}
	\label{tab:table2}
	\begin{tabular}{llll} 
		\hline
		Lens model       & Convergence $\kappa(\theta)$ & Deflection angle $\alpha(\theta)$ & Shear $\gamma(\theta)$ \\
		\hline
        1. Exponential disk & $\displaystyle{\kappa_0{\rm exp}(-{\theta}/{\theta_0})}$ &
                           $\displaystyle{\frac{2\kappa_0}{\theta}\left[ \theta^2_0-\theta_0(\theta+\theta_0){\rm exp}({-{\theta}/{\theta_0}}) \right]}$&
                           $\displaystyle{\frac{\kappa_0}{\theta^2} \left[ 2\theta^2_0-(\theta^2+2\theta\theta_0+2\theta^2_0){\rm exp}({-{\theta}/{\theta_0}}) \right]}$ \\
        2. Gaussian disk    & $\displaystyle{\kappa_0{\rm exp}(-{\theta^2}/{\theta^2_0})}$ &
                           $\displaystyle{\frac{\kappa_0\theta^2_0}{\theta} \left[ 1-{\rm exp}({-{\theta^2}/{\theta^2_0}}) \right]}$ &
                           $\displaystyle{\frac{\kappa_0}{\theta^2} \left[ \theta^2_0- (\theta^2+\theta^2_0){\rm exp}(-{\theta^2}/{\theta^2_0}) \right] }$ \\
        3. Modified Hubble profile & $\displaystyle{\kappa_0{\theta_0^{2}}/{(\theta^{2}+\theta_0^{2})}}$ &
                                  $\displaystyle{\frac{\kappa_0\theta_0^2}{\theta}\ln(\theta^2/\theta_0^2+1) }$ &
                                  $\displaystyle{\frac{\kappa_0\theta_0^{2}}{\theta^2} \left[\ln(\theta^2/\theta_0^2+1)-\frac{\theta^2}{\theta^2+\theta_0^2}\right] }$ \\
		\hline
	\end{tabular}
\end{table*}

\subsection{NFW lens}
As a typical model for describing the density distribution of dark matter halo,
the NFW density profile is written as \citep{1997ApJ...490..493N, 2000ApJ...534...34W} ,
\begin{equation} \label{equ:rhoNFW}
  \rho(r)=\frac{\delta_{\rm c}\rho_{\rm c}}{(r/r_{\rm s})(1+r/r_{\rm s})^2},
\end{equation}
where $\rho_{\rm c}$ is the critical density of the universe. The scale
radius, $r_{\rm s}=r_{200}/c$, is a characteristic radius of the halo,
where $c$ is the concentration parameter
and the characteristic overdensity $\delta_{\rm c}$ for the halo is
\begin{equation}
  \delta_{\rm c}=\frac{200}{3}\frac{c^2}{\ln(1+c)-c/(1+c)}.
\end{equation}

With the thin lens approximation, we define the line of sight as the optical
axis $z$, and the three-dimensional NFW density profile $\rho(D_{\rm d}\bm{\theta}, z)$
will be reduced as two-dimensional surface mass density \citep[e.g., ][]{2002A&A...390..821G, 2014IJAA....4..340H},
\begin{equation}\label{equ:rhoint}
  \Sigma(x) = \int_{-\infty}^{+\infty}\rho(x;z){\rm d}z=2\Sigma_{\rm s}F(x),
\end{equation}
Here the dimensionless radial distance $\bm{x}=\bm{\theta}/\theta_{\rm s}$,
where $\theta_{\rm s}=r_{\rm s}/D_{\rm d}$.
$\Sigma_{\rm s} = \delta_{\rm c}\rho_{\rm c}r_{\rm s}$ is defined as the
characteristic surface mass density, and the factor $F(x)$ is
\begin{equation}
  F(x)=
      \begin{cases}
        \displaystyle{\frac{1}{x^2-1}\left(1-\frac{1}{\sqrt{1-x^2}}\mathrm{arccosh}\frac{1}{x}\right)} & (x<1)\\
        \displaystyle{\frac{1}{3}} & (x=1)\\
        \displaystyle{\frac{1}{x^2-1}\left(1-\frac{1}{\sqrt{x^2-1}}\arccos{\frac{1}{x}}\right)} & (x>1)
      \end{cases}.
\end{equation}
Additionally, the mean surface density inside the dimensionless radius $x$ is
\begin{equation}
  \bar{\Sigma}(x) = \frac{1}{\pi x^2}\int_{0}^{x}2\pi x\Sigma(x){\rm d}x = 4\Sigma_{\rm s}\frac{G(x)}{x^2},
\end{equation}
with
\begin{equation}
  G(x)=
      \begin{cases}
        \displaystyle{\ln{\frac{x}{2}}+\frac{1}{\sqrt{1-x^2}}\mathrm{arcch}\frac{1}{x}} & (x<1)\\
        \displaystyle{\ln{\frac{1}{2}} +1} & (x=1)\\
        \displaystyle{\ln{\frac{x}{2}}+\frac{1}{\sqrt{x^2-1}}\arccos{\frac{1}{x}}} & (x>1)
      \end{cases}.
\end{equation}
Then the deflection angle $\alpha$ between the source and the image, the convergence $\kappa$ and the
shear $\gamma$ can be derived,
\begin{equation} \label{equ:alphaNFW}
  \begin{cases}
    \displaystyle{\alpha(x)=4\kappa_{\rm s}\frac{G(x)}{x}} \\
    \displaystyle{\kappa(x)=2\kappa_{\rm s}F(x)} \\
    \displaystyle{\gamma(x)=\bar{\kappa}(x)-\kappa(x)=2\kappa_{s}\left[\frac{2G(x)}{x^2}-F(x)\right]}
  \end{cases}.
\end{equation}
As discussed in \citet{1996A&A...313..697B}, since (${\rm d}\alpha/{\rm d}x$) is
continuous, NFW lens can still produce three images at most, despite of its central
singularity.

\subsection{Einasto lens}\label{sec:EinastoLens}
Recent $N$-body simulations indicate that a non-singular three-parameter model
such as the Einasto profile can provide a better description of dark matter
haloes in a wide range of halo mass than the NFW profile
\citep[e.g., ][]{2004MNRAS.349.1039N, 2010MNRAS.402...21N, 2007JCAP...07..006E, 2010MNRAS.405..340D, 2016JCAP...01..042S}.
The profile of an Einasto halo is \citep{2011arXiv1108.4905R},
\begin{equation}
  \rho(r)=\rho_{0}{\rm exp}\left[-\left(\frac{r}{h}\right)^{1/n}\right],
\end{equation}
where $\rho_0$ is central density, and $h$ is the scale length, $n$ is the Einasto index.
Clearly, the Einasto profile corresponding to $n = 1$ is an exponential model in 3D, and $n = 1/2$
gives a Gaussian model.
With the thin lens approximation, the projected surface mass density of Einasto lens model
is given by integrating along the line of sight of the 3D density profile as in Eq.~\ref{equ:rhoint},
\begin{equation}
  \Sigma(x)= \int_{-\infty}^{+\infty}\rho(x;z){\rm d}z = 2\int_{x}^{\infty}\frac{\rho(r)r{\rm d}r}{\sqrt{r^{2}-x^{2}}},
\end{equation}
where $r=\sqrt{x^{2}+z^{2}}$, and we has rewritten the integration as Abel transform
\citep{1987gady.book.....B} in the right term. 
As studied in \citet{2012A&A...546A..32R, 2012A&A...540A..70R},
they derived the surface mass density of Einasto by a Mellin integral transform
formalism, and related lensing properties in terms of the Fox $H$ and Meijer $G$
functions. More details can be referred to \citet{2012A&A...540A..70R}.


Here, as a comparison with the exponential disk lens and Gaussian disk lens model in 2D,
we will present the lensing properties for the two often used models, exponential
model and Gaussian model, in 3D.
Using the specific Meijer $G$ function, we can derive the surface mass density profile 
and lensing properties. Tab.~\ref{tab:tableEinastoLensing} shows the lensing properties 
for the two considered models, and it is found that Einasto model with $n=1/2$ can be
reduced to a Gaussian disk lens model.

\begin{table*}
  \renewcommand\arraystretch{1.5}
  \centering
  \caption{The lensing functions $\alpha(\theta)$, $\kappa(\theta)$, and $\gamma(\theta)$
           for Einasto model with $n=1$ and $n=1/2$. Here $\kappa_{s}$ is the central
           convergence, and $x$ defines the dimensionless radius as $x=\theta/\theta_{\rm s}$.
		   $K_{\nu}(x)$ is the modified Bessel function of the second kind of order $\nu$.}
  \label{tab:tableEinastoLensing}
  \begin{tabular}{llll}
	\hline
    Einasto index $n$ & Convergence $\kappa(x)$ & Deflection angle $\alpha(x)$ & Shear $\gamma(x)$ \\
    \hline
    $n=1$   & $\displaystyle{\kappa_{\rm s} xK_1(x)}$  &
              $\displaystyle{\frac{4\kappa_{\rm s}}{x}\left[1-\frac{x^2}{2}K_2(x)\right]}$ &
              $\displaystyle{\frac{4\kappa_{\rm s}}{x^2}\left[1-\frac{x^2}{2}K_2(x)-\frac{x^3}{4}K_1(x)\right]}$  \\
    $n=1/2$ & $\displaystyle{\kappa_{\rm s} {\rm exp}(-x^2)}$ &
              $\displaystyle{\frac{\kappa_{\rm s}}{x}\left[1-{\rm exp}(-x^2)\right]}$ &
              $\displaystyle{\frac{\kappa_{\rm s}}{x^2}\left[1-\left(1+x^2\right){\rm exp}(-x^2)\right]}$ \\
	\hline
  \end{tabular}
\end{table*}

\section{Magnification invariants}\label{sec:method}
\subsection{The Test of Magnification Summation}
With the purpose of calculating the accurate position of the image and
its corresponding magnification for a given source, we utilize the analytical
ray-tracing method in different image regions to numerically evaluate
magnification for each image. This precise numerical method has been
introduced in \citet{2016MNRAS.461.4466C} to study the magnification relations
of quad lenses, which is shortly described as follows.

\begin{figure}
	\includegraphics[width=0.65\columnwidth]{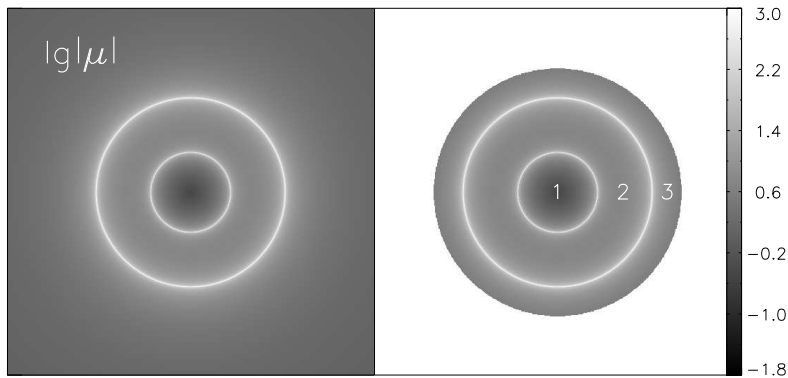}
	\centering
    \caption{The distribution of different image regions by a circularly-symmetric lens.
             The left panel shows the absolute value of the magnification $\mu$ in logarithm.
             The right panel shows the three image regions corresponding to the source region
             inside of the caustic, where  the gray values are derived from the left panel.}
    \label{fig:IMAGEZONE}
\end{figure}

For a well-defined circular lens model, when the source lies in the radial
caustic, one can find three images at most. Therefore, we pixelate the region
in the radial caustic as the source area, so as to find images of it by the
analytical deflection angles. As shown in Fig.~\ref{fig:IMAGEZONE}, the image
of the source area can be divided into three parts in the lens plane. Each
point source in the source region has three corresponding images. The saddle
image lies in image region 2 with negative magnification. The other two images,
lying in image region 1 and 3, arise at the maxima and minima of the time delay
surface, hence they have positive magnification.

For the sake of calculating the magnifications of the three images for a given
point source $\mathcal{P}$ inside of the caustic, we need to obtain the exact
positions of the three images in the lens plane. At first, we set a bundle of
light rays from the observer to grids in the image region 1, and then the deflected
light rays will be traced back to the source region. Clearly, the image position
of the source $\mathcal{P}$ can be approximately estimated by the nearest light
ray to the given point source. Starting from this approximate position of image,
we are able to calculate the image positions accurately for each point source
in the source region by using Newton-Raphson method.

After deriving the positions of the images and theirs magnifications $\mu$ for
the given lens model, we can map the divided image regions in right panel of
Fig.~\ref{fig:IMAGEZONE} to the source plane, and estimate the magnification summation
for each source. Using this numerical method, images can be matched to their source
precisely in different image region.

\begin{figure*}
	\centering
	\includegraphics[width=0.95\columnwidth]{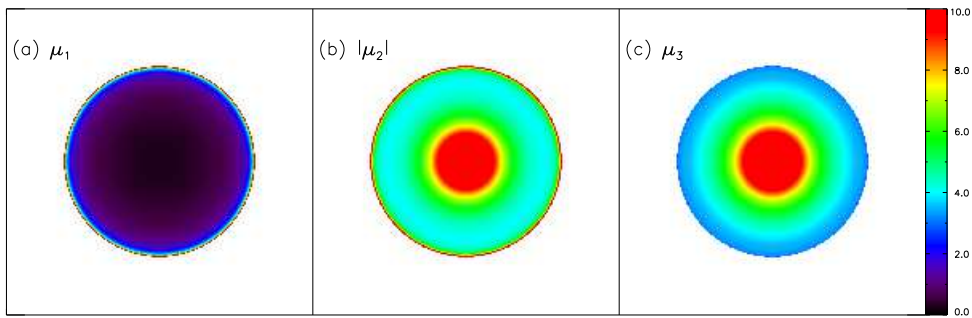}
    \caption{The magnification distributions as functions of source positions.
             The values are mapped from three different image regions to source
             plane respectively. The left panel shows magnification $\mu_{1}$ of
             the image region 1, which is corresponding to the maxima of the time
             delay surface. The middle panel presents images that arise at the
             saddle points of the time delay surface with the negative magnification
             $\mu_{2}$, so here we flipped the sign of the magnification. The
             magnification $\mu_{3}$ of image region 3, which arises at the minima
             of the time delay surface, is shown in the right panel.}
    \label{fig:expFlux}
\end{figure*}

For the exponential disk lens, we have analytically derived the scaled deflection angle
$\alpha$ and the magnification $\mu$ from the 2D Poisson equation in section \ref{ssec:simpleLens}.
Employing the numerical method introduced above, we can map the magnifications of the
three different regions in right panel of Fig.~\ref{fig:IMAGEZONE} into the source plane
accurately, and the results are shown in Fig.~\ref{fig:expFlux}.
The left panel shows magnification $\mu_{1}$ of the image region 1, which is corresponding
to the maxima of the time delay surface, and generally demagnified. The middle panel presents
images that arise at the saddle points of the time delay surface with the negative
magnification $\mu_{2}$. Here we flipped the sign of the magnification. The magnification
$\mu_3$ of image region 3, which arises at the minima of the time delay surface, is shown in
the right panel.

\begin{figure}[h]
  \begin{minipage}[t]{0.495\linewidth}
  \centering
   \includegraphics[width=75mm]{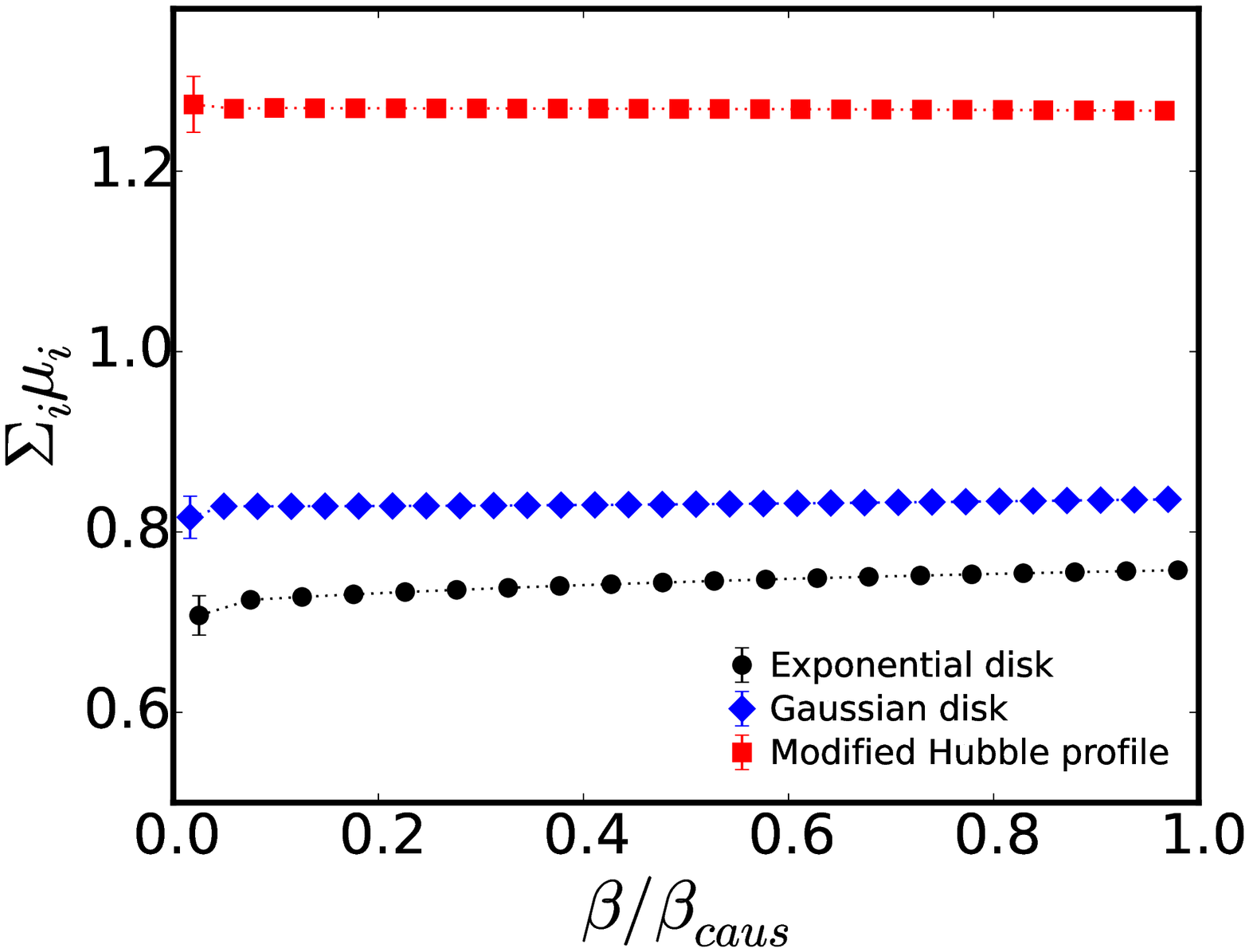}
  \end{minipage}%
  \begin{minipage}[t]{0.495\textwidth}
  \centering
   \includegraphics[width=75mm]{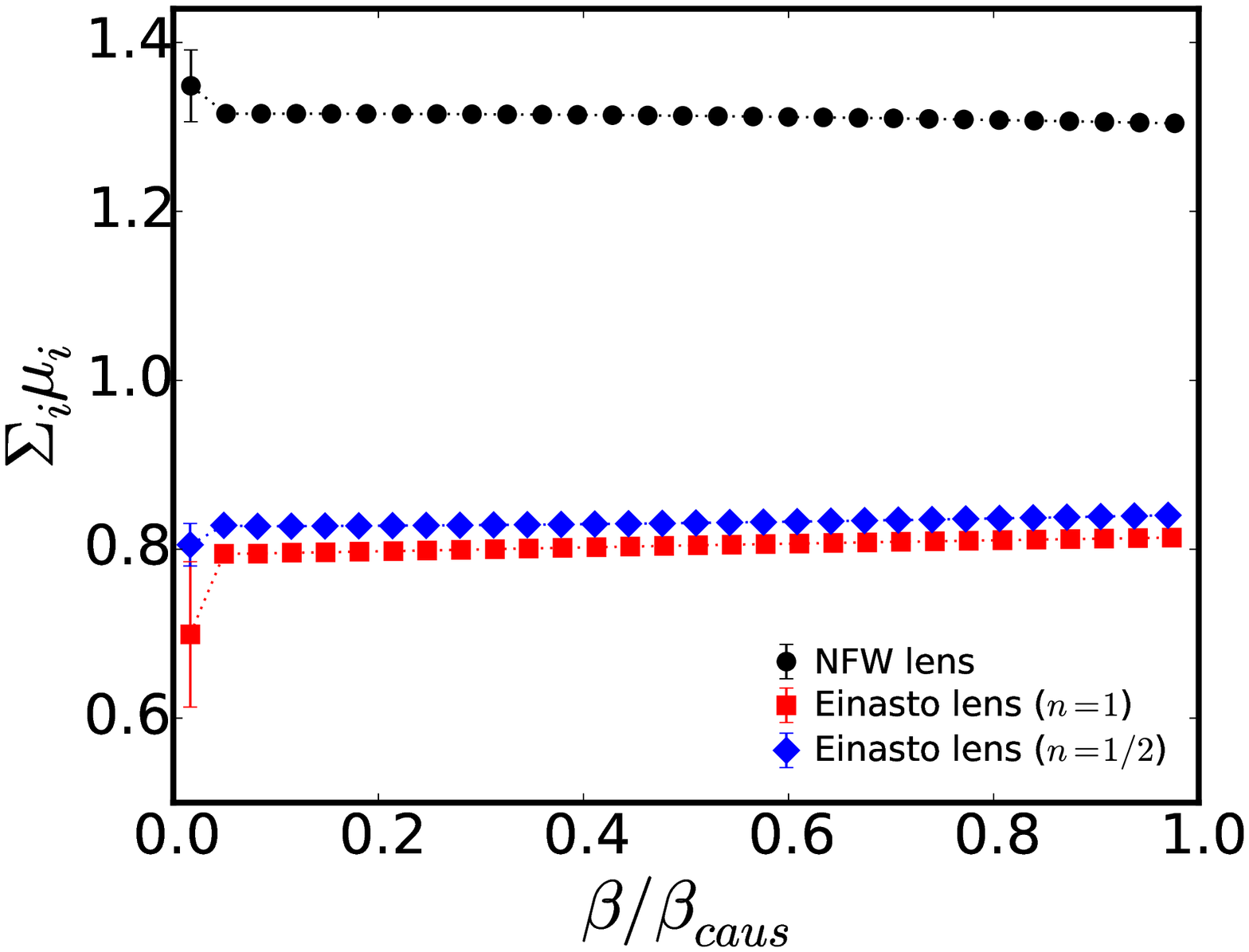}
  \end{minipage}%
  \caption{The summation of signed magnifications of images for the given
           disk lens models in the left panel and the NFW/Einasto lens models
           in the right panel. Source positions are scaled by
           the radius of radial caustic, $\beta_{{\rm caus}}$.}
  \label{fig:FluxSum}
\end{figure}
With these image magnification in different image region shown in Fig.~\ref{fig:expFlux},
we calculate the magnification summation by $\Sigma_{i}\mu_{i}=\mu_{1}+\mu_{2}+\mu_{3}$,
and the result is shown in the left panel of Fig.~\ref{fig:FluxSum} by black dots.
Here the source position $\beta$ has been normalized by the radial caustic $\beta_{\rm caus}$,
and the error bars are estimated from the different radial bins. Note that the large
error around the center happens because of the magnification changes fast with the
increasing of the radius.
For arbitrary source positions within the caustic for a given set of $\kappa_{0}$ and
$\theta_{0}$, the summed magnification is almost constant, but with a subtle change,
and can be evaluated by the mean value as $0.75{\pm}0.03$ here.

Analogous to exponential disk lens model, we also calculate magnification summation
for Gaussian model and modified Hubble profile lens model as shown in the left
panel of Fig.~\ref{fig:FluxSum}, and in the right panel we show the magnification
summations for the NFW lens and Einasto lens. We find the magnification summations
are almost constant for these lens models, regardless of the position for the source
in the radial caustic.

\subsection{Dependence of Invariants on the Model Parameters}\label{sec:evo4param}
In general, for a two-parameter circular lens model, its dimensionless surface
mass density can be generally defined by the characteristic surface mass  density
$\kappa_{{\rm c}}$ and the radial distribution function $f(\theta/\theta_{\rm c})$,
\begin{equation}
  \kappa(\theta) = \kappa_{{\rm c}} f(\theta/\theta_{\rm c}),
\end{equation}
where $\kappa_{{\rm c}}$ can be $\kappa_{0}$ in the disk lens model or $\kappa_{\rm s}$
in NFW and Einasto lens models,
$\theta_{{\rm c}}$ is the characteristic radius of the lens model, such as $\theta_{0}$
in exponential disk lens model and $\theta_{\rm s}$ in NFW and Einasto lens models.
If a dimensionless radial distance $x=\theta/\theta_{\rm c}$ is defined in these circular
lens models, we find  the lensing properties will not be changed. Thus the magnification
invariants should not depend on the characteristic radius $\theta_{{\rm c}}$.
For given model parameters ($\kappa_{{\rm c}}, \theta_{{\rm c}}$), the magnification invariant can be
numerically estimated as represented above. While the dependency between the invariant
and the model parameters is unclear. In this section, we test how the magnification
invariant depends on the models parameters $\kappa_{{\rm c}}$ and $\theta_{{\rm c}}$.

\begin{figure*}
  \centering
  \includegraphics[width=0.98\columnwidth]{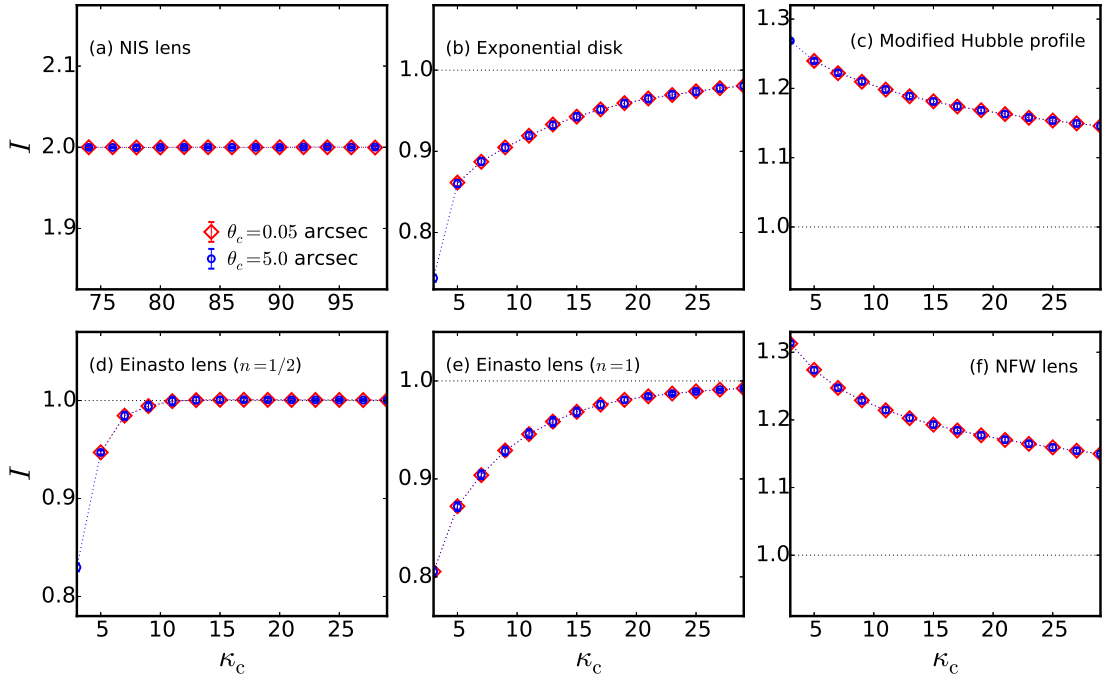}
  \caption{The magnification invariants with the different model parameters
           $\kappa_{{\rm c}}$ and $\theta_{{\rm c}}$. NIS model shows a constant
           magnification invariant, which is independent of model parameters.
           While, for the other lens models, the characteristic surface mass
           density $\kappa_{{\rm c}}$ can change the invariants significantly.}
  \label{fig:test-k0}
\end{figure*}

We first verify the feasibility of our method for determining magnification invariant by
considering the well defined lens model with the non-singular isothermal sphere, termed
NIS \citep[e.g., ][]{1994A&A...284..285K, 2007MNRAS.376..113A}, which is an isothermal
sphere with small but finite core,
\begin{equation}
  \kappa_{{\rm NIS}}(\theta) = \frac{\kappa_{\rm c}}{2\sqrt{\theta^2/\theta_{{\rm c}}^2 + 1}}.
\end{equation}
For this specific model of the circular lens, one can analytically calculate the magnification
summation of the three images. In Appendix~\ref{appendixA}, we theoretically prove that the
magnification invariant of NIS equals $2$, which is a constant without dependence of models
parameters $\kappa_{{\rm c}}$ and $\theta_{{\rm c}}$. As a comparison, we estimate the
magnification invariant $I$ through our numerical method for the NIS model.
The invariants are evaluated by the mean value of the magnification summation $\sum_{i=1}^{3}\mu_{i}$
at different source position, as the function of different model parameters, $\kappa_{{\rm c}}$
and $\theta_{{\rm c}}$. Fig.~\ref{fig:test-k0}(a) shows the magnification invariants with different
parameters for NIS model. Clearly, our numerical result indicates that the invariant of NIS lens
is equal to 2, and this magnification invariant is independent of model parameters, which agrees
well with the prediction in Appendix~\ref{appendixA}.

Using our numerical method, we investigate magnification invariants for the given
lens models, and show the results in  Fig.~\ref{fig:test-k0}. We find that the invariants do not
change for different scale length $\theta_{{\rm c}}$ as we expected. But the characteristic surface mass density
$\kappa_{{\rm c}}$ can change the invariants significantly. When $\kappa_{{\rm c}}$ is large enough,
the magnification invariants converge to 1 for both exponential disk lens and Einasto lens. 
Note that for Einasto model the index $n$ is very important in determining 
the surface mass density characteristics, which determine the lensing properties  
of the respective profiles. Thus with different index $n$, the converged 
value of magnification invariants may be different. 
Moreover,
the invariants of exponential disk lens and Einasto lens (both $n=1/2$ and $1$) increase monotonically
over the characteristic surface mass density $\kappa_{{\rm c}}$, while decrease for the modified
Hubble profile and NFW lens. 

\begin{figure*}
  \centering
  \includegraphics[width=0.65\columnwidth]{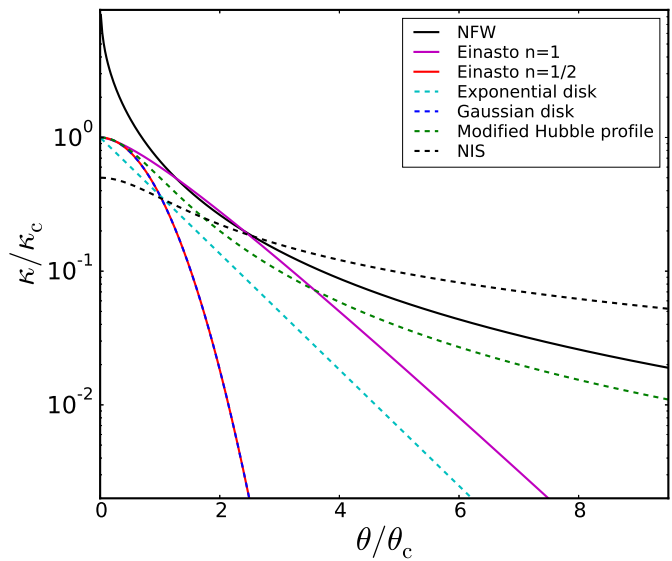}
  \caption{Density profiles of the given lens models.}
  \label{fig:profile}
\end{figure*}

This difference should be resulted by the different density profiles of the lens models. 
Fig.~\ref{fig:profile} shows the density profile of the used lens models in this paper.
Clearly, for the exponential disk lens and Einasto lens (both $n=1/2$ and $1$), their density profiles  
remarkably decrease to $\lesssim 10^{-2}\kappa_{\rm c}$ at $\theta \sim 5\theta_{\rm c}$. While for NFW 
and modified Hubble model, the density profile decrease gradually at the outer part ($\theta > 5\theta_{\rm c}$) 
of the lens model. For the NIS model, its density profile decreases with a more gentle slope, 
and this lens model gives a higher magnification invariant, which equals 2. Thus in other words the slope of 
density profile can be a potential explanation for the different converged value of magnification invariants 
between the different lens models. 
Moreover, as indicated from the density profile, the total mass is convergence for the exponential
disk lens and Einasto lens, but it is divergence for the lens with modified Hubble model and NFW lens.
One of another potential explanation is the ratio of mass inside of the Einstein radius to
the total mass, $F=M(\leq \theta_{\rm E})/M_{\rm tot}$. We find that, for both exponential disk
lens and Einasto lens ($n=1/2$ and $1$), this mass ratio tends to $1$ with the increasing of the
characteristic surface mass density $\kappa_{\rm c}$, while the mass ratio $F \rightarrow 0$ for the modified
Hubble profile and NFW lens. Combining with that $F = 1$ for the point lens and $F = 0$ for SIS and NIS,
we speculate that when $F \rightarrow 1$ of a given lens, it will act as a point lens and have a magnification
invariant of $I=1$. While a lens with $F \rightarrow 0$, for which the total mass is not converge, can hardly
have a magnification invariant of $I \rightarrow 1$ with the increasing of its model parameters. More detailed
analyses in the future should be helpful to understand the contrast between these two different results.

\section{CONCLUSION AND DISCUSSION}\label{sec:summary}
Forming multiple images of the given background source is one of the most important effects of
strong gravitational lensing. When the image number is a maximum, the summation of signed
magnifications of the images can be a constant for certain lens models. For some quadruple
lens models, it is found that the invariant is independent of most of the model parameters, 
as long as the source lies inside of the caustic \citep{1998ApJ...509L..13D}.
In this paper, we focus on several commonly used circular lens models, in which the magnification
invariants have not been investigated before.
We find that for the exponential disk lens model, magnification invariant should be existent,
but with a very subtle change as changing of source position $\beta$ in the caustic region. Moreover, we also
calculate the magnifications of the other two important lens models, Gaussian disk lens and modified
Hubble profile lens. Our results indicate that magnification invariants do also exist for these two lens models.

Considering the dark matter halo can be described by some universal profiles, we further examine two
typical models: singular two-parameter model, e.g. NFW lens, and non-singular three-parameter model,
e.g. Einasto lens. With the thin lens approximation, we find that the magnification summations
of three images for an arbitrary point source inside the caustic are constant for both the
NFW and Einasto models .

More tests indicate that the magnification summation does not change with changing of the scale length
$\theta_{\rm c}$, because of that the lensing properties will not be changed if a dimensionless radial
distance $x=\theta/\theta_{\rm c}$ is defined in the circular lens models. Thus the magnification
invariants are independent of the characteristic radius $\theta_{{\rm c}}$.
However, the central density $\kappa_{\rm c}$ can affect the magnification summations
significantly. In the paper, we show how the magnification summations of a given
lens model vary as functions of the model parameters $\kappa_{\rm c}$ and $\theta_{\rm c}$. When
$\kappa_{{\rm c}}$ is very large, the magnification invariant tends to be 1 for both exponential disk
lens and Einasto lens (both $n=1/2$ and $1$), which can be explained as that the lens will act as a point
mass with $\kappa_{{\rm c}}$ increasing. Moreover, the invariants of exponential disk lens and Einasto
lens increase monotonically with the characteristic surface mass density $\kappa_{{\rm c}}$, while it is
decreasing for the modified Hubble profile and NFW lens. This difference should be resulted by the different
density profiles of the lens models, and more detailed analyses  should be helpful to understand the contrast
between these two different results.

Observationally, combining the magnification invariants and the measurements of Type Ia supernovae (SNe Ia),
which are excellent standard candles, we can constrain the lens model in more detail. While the invariant
can be significantly affected by the substructure or high-order asymmetry of lens. In the future, the
magnification of high-order lens model and of the haloes in $N$-body simulation or hydrodynamic simulation
should be studied, and the magnification invariant can be examined in a more realistic manner.

\begin{acknowledgements}
The authors thank Qianli Xia for helpful
discussion and suggestions. We acknowledge the
supports of the NSFC (No. 11403103, 11603032, 11333008, 11273061), 
the 973 program (No. 2015CB857003, 2013CB834900), 
China Postdoctoral Science Foundation (2014M551681) 
and the NSF of Jiangsu province (No. BK20140050).
\end{acknowledgements}

\appendix                  
\section{Magnification invariant for NIS}\label{appendixA}
As an important lens model, the Singular Isothermal Sphere (SIS) has been widely used in the theoretical reserch.
Another one more realistic model for modeling lensing galaxies is Non-singular Isothermal Sphere (NIS), which is profiled by isothermal sphere with a finite core $\theta_{{\rm c}}$.
In this case, by defining dimensionless radius $\bm{x} = \bm{\theta}/\theta_{{\rm c}}$, the convergence of NIS is given by
\begin{equation}
  \kappa_{{\rm NIS}}(x) = \frac{\kappa_{{\rm c}}}{2\sqrt{x^2+1}} ,
\end{equation}
where $\kappa_{{\rm c}}$ is a constant for the given radial profile.
Clearly, using 2D Poisson equation, the lensing functions can be derived as:
\begin{equation}
  \alpha_{{\rm NIS}}(x) = \frac{\kappa_{{\rm c}}}{x}\left( \sqrt{x^2+1}-1 \right),
\end{equation}
\begin{equation}
  \gamma_{{\rm NIS}}(x) = \frac{\kappa_{{\rm c}}}{x^2}\left( \sqrt{x^2+1}-1 \right) - \frac{\kappa_{{\rm c}}}{2\sqrt{x^2+1}}.
\end{equation}
Then, for a given source position $y_{0}$, combining the lens equation $y_{0} = x - \alpha_{{\rm NIS}}(x)$ and the magnification function $1/\mu(x) = \left[ 1-\kappa_{{\rm NIS}}(x) \right]^2-\gamma_{{\rm NIS}}(x)^2$, we can eliminate $x$ to write a 3rd-degree polynomial equations about $\mu$ as

\begin{equation}
A\mu^3+B\mu^2+C\mu+D=0 .
\end{equation}
Here the four coefficients related to the source position $y_{0}$ and the model parameter $\kappa_{{\rm c}}$ are,
\begin{equation*}
  \begin{cases}
    \displaystyle{A = y_{0}^{2} \left[ y_{0}^{4}-y_{0}^{2}(2\kappa_{{\rm c}}^{2}+10\kappa_{{\rm c}}-1)+\kappa_{{\rm c}}(\kappa_{{\rm c}}-2)^3 \right]}, \\
    \displaystyle{B = -2 y_{0}^{2} \left[ y_{0}^{4}-y_{0}^{2}(2\kappa_{{\rm c}}^{2}+10\kappa_{{\rm c}}-1)+\kappa_{{\rm c}}(\kappa_{{\rm c}}-2)^3 \right]}, \\
    \displaystyle{C = y_{0}^{6}-y_{0}^{4}(3\kappa_{{\rm c}}^{2}+10\kappa_{{\rm c}}-1) +y_{0}^{2}\kappa_{{\rm c}}(3\kappa_{{\rm c}}^{3}+4\kappa_{{\rm c}}^{2} }\\\displaystyle{\qquad +8\kappa_{{\rm c}}-8)-\kappa_{{\rm c}}^{2}(\kappa_{{\rm c}}^{2}-3\kappa_{{\rm c}}+2)^{2} }, \\
    \displaystyle{D = 4\kappa_{{\rm c}}^2 \left[ y_{0}^{2}+(\kappa_{{\rm c}}-1)^{2} \right]}. \\
  \end{cases}
\end{equation*}
Thus, we can obtain three roots at most, and these three magnifications ($\mu_{1}$, $\mu_{2}$, $\mu_{3}$) correspond to the three images of source at $y_{0}$ for the given NIS parameter $\kappa_{{\rm c}}$.
As described by the Vieta's formulas in mathematics, we are able to calculate the sum of these magnifications by
\begin{equation}
  \sum_{i=1}^{3}\mu_{i} = -\frac{B}{A} = 2.
\end{equation}
Therefore, one can rigidly prove that the magnification invariant is $I = \sum_{i}\mu_{i} = 2$ for the NIS lens model, providing the source is inside the caustic. Moreover, this invariant is independent of the parameters $y_{0}$ and $\kappa_{{\rm c}}$.


\label{lastpage}

\begin{thebibliography}{99}

\bibitem[Aazami \& Petters(2009)]{2009JMP....50c2501A} Aazami, A.~B., \& Petters, A.~O.\ 2009, Journal of Mathematical Physics, 50, 032501 
\bibitem[Aubert et al.(2007)]{2007MNRAS.376..113A} Aubert, D., Amara, A., \& Metcalf, R.~B.\ 2007, \mnras, 376, 113 
\bibitem[Bartelmann(1996)]{1996A&A...313..697B} Bartelmann, M.\ 1996, \aap, 313, 697 
\bibitem[Biggs et al.(1999)]{1999MNRAS.304..349B} Biggs, A.~D., Browne, I.~W.~A., Helbig, P., et al.\ 1999, \mnras, 304, 349 
\bibitem[Binney \& Tremaine(1987)]{1987gady.book.....B} Binney, J., \& Tremaine, S.\ 1987, Princeton, NJ, Princeton University Press, 1987
\bibitem[Burke(1981)]{1981ApJ...244L...1B} Burke, W.~L.\ 1981, \apjl, 244, L1 
\bibitem[Cao et al.(2015)]{2015ApJ...806..185C} Cao, S., Biesiada, M., Gavazzi, R., Pi{\'o}rkowska, A., \& Zhu, Z.-H.\ 2015, \apj, 806, 185
\bibitem[Chu et al.(2015)]{2015MNRAS.449.2079C} Chu, Z., Li, G.~L., \& Lin, W.~P.\ 2015, \mnras, 449, 2079 
\bibitem[Chu et al.(2016)]{2016MNRAS.461.4466C} Chu, Z., Li, G.~L., Lin, W.~P., \& Pan, H.~X.\ 2016, \mnras, 461, 4466 
\bibitem[Dalal(1998)]{1998ApJ...509L..13D} Dalal, N.\ 1998, \apjl, 509, L13 
\bibitem[Dalal \& Rabin(2001)]{2001JMP....42.1818D} Dalal, N., \& Rabin, J.~M.\ 2001, Journal of Mathematical Physics, 42, 1818 
\bibitem[Dhar \& Williams(2010)]{2010MNRAS.405..340D} Dhar, B.~K., \& Williams, L.~L.~R.\ 2010, \mnras, 405, 340 
\bibitem[Dyer \& Roeder(1980)]{1980ApJ...238L..67D} Dyer, C.~C., \& Roeder, R.~C.\ 1980, \apjl, 238, L67 
\bibitem[El{\'{\i}}asd{\'o}ttir \& M{\"o}ller(2007)]{2007JCAP...07..006E} El{\'{\i}}asd{\'o}ttir, {\'A}., \& M{\"o}ller, O.\ 2007, \jcap, 7, 006 
\bibitem[Fassnacht et al.(2002)]{2002ApJ...581..823F} Fassnacht, C.~D., Xanthopoulos, E., Koopmans, L.~V.~E., \& Rusin, D.\ 2002, \apj, 581, 823 
\bibitem[Golse \& Kneib(2002)]{2002A&A...390..821G} Golse, G., \& Kneib, J.-P.\ 2002, \aap, 390, 821 
\bibitem[Hurtado et al.(2014)]{2014IJAA....4..340H} Hurtado, R., Casta{\~n}eda, L., \& Tejeiro, J.~M.\ 2014, International Journal of Astronomy and Astrophysics, 4, 340 
\bibitem[Koopmans et al.(2009)]{2009ApJ...703L..51K} Koopmans, L.~V.~E., Bolton, A., Treu, T., et al.\ 2009, \apjl, 703, L51 
\bibitem[Kormann et al.(1994)]{1994A&A...284..285K} Kormann, R., Schneider, P., \& Bartelmann, M.\ 1994, \aap, 284, 285 
\bibitem[Leh{\'a}r et al.(2000)]{2000ApJ...536..584L} Leh{\'a}r, J., Falco, E.~E., Kochanek, C.~S., et al.\ 2000, \apj, 536, 584 
\bibitem[Linder(2011)]{2011PhRvD..84l3529L} Linder, E.~V.\ 2011, \prd, 84, 123529 
\bibitem[Mao et al.(2001)]{2001MNRAS.323..301M} Mao, S., Witt, H.~J., \& Koopmans, L.~V.~E.\ 2001, \mnras, 323, 301 
\bibitem[McKean et al.(2015)]{2015aska.confE..84M} McKean, J., Jackson, N., Vegetti, S., et al.\ 2015, Advancing Astrophysics with the Square Kilometre Array (AASKA14), 84 
\bibitem[Meylan et al.(2006)]{2006glsw.conf.....M} Meylan, G., Jetzer, P., North, P., et al.\ 2006, Saas-Fee Advanced Course 33: Gravitational Lensing: Strong, Weak and Micro (Berlin: Springer)
\bibitem[Miralda-Escude(1991)]{1991ApJ...370....1M} Miralda-Escude, J.\ 1991, \apj, 370, 1 
\bibitem[Narayan \& Bartelmann(1999)]{1999fsu..conf..360N} Narayan, R., \& Bartelmann, M.\ 1999, Formation of Structure in the Universe, 360 
\bibitem[Navarro et al.(2004)]{2004MNRAS.349.1039N} Navarro, J.~F., Hayashi, E., Power, C., et al.\ 2004, \mnras, 349, 1039 
\bibitem[Navarro et al.(1997)]{1997ApJ...490..493N} Navarro, J.~F., Frenk, C.~S., \& White, S.~D.~M.\ 1997, \apj, 490, 493 
\bibitem[Navarro et al.(2010)]{2010MNRAS.402...21N} Navarro, J.~F., Ludlow, A., Springel, V., et al.\ 2010, \mnras, 402, 21 
\bibitem[Paraficz(2009)]{2009PhDT.......238P} Paraficz, D.\ 2009, Ph.D.~Thesis
\bibitem[Petters \& Werner(2010)]{2010GReGr..42.2011P} Petters, A.~O., \& Werner, M.~C.\ 2010, General Relativity and Gravitation, 42, 2011
\bibitem[Retana-Montenegro \& Frutos-Alfaro(2011)]{2011arXiv1108.4905R} Retana-Montenegro, E., \& Frutos-Alfaro, F.\ 2011, arXiv:1108.4905 
\bibitem[Retana-Montenegro et al.(2012a)]{2012A&A...546A..32R} Retana-Montenegro, E., Frutos-Alfaro, F., \& Baes, M.\ 2012, \aap, 546, A32 
\bibitem[Retana-Montenegro et al.(2012b)]{2012A&A...540A..70R} Retana-Montenegro, E., van Hese, E., Gentile, G., Baes, M., \& Frutos-Alfaro, F.\ 2012, \aap, 540, A70 
\bibitem[Rood et al.(1972)]{1972ApJ...175..627R} Rood, H.~J., Page, T.~L., Kintner, E.~C., \& King, I.~R.\ 1972, \apj, 175, 627 
\bibitem[Rusin et al.(2005)]{2005ApJ...627L..93R} Rusin, D., Keeton, C.~R., \& Winn, J.~N.\ 2005, \apjl, 627, L93 
\bibitem[Schneider et al.(1992)]{1992grle.book.....S} Schneider, P., Ehlers, J., \& Falco, E.~E.\ 1992, Gravitational Lenses, XIV, (Berlin: Springer), 112 
\bibitem[Sereno et al.(2016)]{2016JCAP...01..042S} Sereno, M., Fedeli, C., \& Moscardini, L.\ 2016, \jcap, 1, 042 
\bibitem[Shu et al.(2016)]{2016ApJ...833..264S} Shu, Y., Bolton, A.~S., Mao, S., et al.\ 2016, \apj, 833, 264 
\bibitem[Shu et al.(2017)]{2017ApJ...851...48S} Shu, Y., Brownstein, J.~R., Bolton, A.~S., et al.\ 2017, \apj, 851, 48 
\bibitem[Suyu et al.(2014)]{2014ApJ...788L..35S} Suyu, S.~H., Treu, T., Hilbert, S., et al.\ 2014, \apjl, 788, L35 
\bibitem[Tortora(2007)]{2007waas.work..127T} Tortora, C.\ 2007, 1st Workshop of Astronomy and Astrophysics for Students, 127 
\bibitem[Trotter et al.(2000)]{2000ApJ...535..671T} Trotter, C.~S., Winn, J.~N., \& Hewitt, J.~N.\ 2000, \apj, 535, 671 
\bibitem[Tsukamoto \& Harada(2013)]{2013PhRvD..87b4024T} Tsukamoto, N., \& Harada, T.\ 2013, \prd, 87, 024024 
\bibitem[Vegetti et al.(2012)]{2012Natur.481..341V} Vegetti, S., Lagattuta, D.~J., McKean, J.~P., et al.\ 2012, \nat, 481, 341 
\bibitem[Werner(2009)]{2009JMP....50h2504W} Werner, M.~C.\ 2009, Journal of Mathematical Physics, 50, 082504 
\bibitem[Witt \& Mao(2000)]{2000MNRAS.311..689W} Witt, H.~J., \& Mao, S.\ 2000, \mnras, 311, 689 
\bibitem[Wright \& Brainerd(2000)]{2000ApJ...534...34W} Wright, C.~O., \& Brainerd, T.~G.\ 2000, \apj, 534, 34 
\bibitem[Yuan \& Wang(2015)]{2015MNRAS.452.2423Y} Yuan, C.~C., \& Wang, F.~Y.\ 2015, \mnras, 452, 2423 


\end{thebibliography}
\end{document}